# Probability Distribution of the Hubble Constant and the Age of the Universe Inferred from the Local Observations

Takahiro Nakamura and Yasushi Suto

*Department of Physics, The University of Tokyo, Tokyo 113, Japan.*
e-mail: nakamura@yayoi.phys.s.u-tokyo.ac.jp and suto@phys.s.u-tokyo.ac.jp

**Abstract**

We present a method to compute the probability distribution function of the (true) Hubble constant and the age of the universe, given the estimate of the Hubble constant in our nearby galaxy samples. Our method takes into account both the observational errors and the cosmic variance, and enables to quantitatively compute the constraints on the cosmological models. Based on the present local observation $H_0 = 80 \pm 17$ km/s/Mpc, the probability of $H_0 < 50$ km/s/Mpc is about 6% for the Einstein – de Sitter universe ($\Omega_0 = 1$) and 4% for an open ($\Omega_0 = 0.2$) universe. These probabilities are reduced to 0.8% and 0.03%, respectively, if the accuracy of the observational uncertainty is improved within 10%. Similar probabilistic constraints on $t_0$ are also discussed.

## 1. Introduction

The Hubble constant $H_0(\equiv 100h$ km/s/Mpc) is one of the fundamental parameters in cosmology which fixes the scale of the various important quantities in astrophysics. The recent observations of the Cepheid variables in the Virgo cluster by Hubble Space Telescope suggested a somewhat large value, $h = 0.8 \pm 0.17$ (Freedman et al. 1994). On the other hand, the Sunyaev-Zel'dovich effect and the time delay due to the gravitational lens effect predict smaller values around $h \sim 0.5$ (e.g., Birkinshaw & Hughes 1994; Yamashita 1994; Gorenstein, Falco, & Shapiro 1988; Rhee 1991). Although the latter methods contain considerable uncertainties, these estimates may be close to the true (global) value because these are the high redshift observations. In general, lower $h$ is desirable in order to reconcile the age of the universe $t_0 \sim H_0^{-1}$ with that of the oldest stars in the globular cluster (e.g., Chaboyer et al. 1992).

It is possible that the Hubble constant determined from the local observations systematically differs from its global value due to the density inhomogeneity on the scale of the observational samples; the expansion rate of small patches of the universe varies from place to place depending on the degree of such inhomogeneities (Suto, Suginohara, & Inagaki 1995). In addition, the uncertainties intrinsic to the local observations complicate the estimates of the true value of $H_0$. To what extent can we estimate $H_0$ and $t_0$ on the basis of the value derived from the local observations ? To quantitatively discuss the cosmological implications from the local Hubble constant estimate, it is important to derive the probability distribution function (PDF) of the global Hubble constant.

Previously, Turner, Cen, & Ostriker (1992, hereafter TCO) considered the difference between the local and global Hubble constant directly from the N-body simulation data, and calculated the PDF numerically. Suto,

Suginohara, & Inagaki (1995) derived an analytical expression for the lower limit on $H_0$ and the upper limit on $t_0$ with given value of the local Hubble constant.

In this *Letter*, we present a method to compute the PDF of the global Hubble constant using the non-linear spherical model. Our semi-analytical expression applied to the cold dark matter (CDM) model reproduces the numerical result by TCO very well. We discuss the constraints on the cosmological parameters quantitatively in the CDM model as an example.

## 2. Matching the Local and Global Universes in a Spherical Approximation

Let us consider the matching of the homogeneous global universe with the local spherical region of radius $r$ and mass $M$, first in the case of a vanishing cosmological constant $\Lambda$. For open universes, the expansion equations are given parametrically as (e.g., eqs.[19.11] and [97.21] in Peebles 1980)

$$r = \frac{GM}{2E}(\cosh\theta - 1), \qquad t = \frac{GM}{(2E)^{3/2}}(\sinh\theta - \theta), \tag{1}$$

for the local spherical region, and

$$a = \frac{GM}{-K}(\cosh\eta - 1), \qquad t = \frac{GM}{(-K)^{3/2}}(\sinh\eta - \eta), \tag{2}$$

for the global universe, where $E$ and $-K$ are some positive constants. For the later discussion, we define the following conventional variables with the subscript $L$ denoting those in the local region:

$$H_L \equiv \frac{\dot{r}}{r}, \quad \Omega_L \equiv \frac{8\pi G\rho_L}{3H_L^2}, \quad \rho_L \equiv \frac{3M}{4\pi r^3}, \tag{3}$$

$$H \equiv \frac{\dot{a}}{a}, \quad \Omega \equiv \frac{8\pi G\rho}{3H^2}, \quad \rho \equiv \frac{3M}{4\pi a^3}, \tag{4}$$

$$\delta \equiv \frac{\rho_L - \rho}{\rho}, \qquad \delta_H \equiv \frac{H_L - H}{H}. \tag{5}$$

With equations (1) and (2), a straightforward calculation yields

$$\delta = \left(\frac{\sinh\theta - \theta}{\sinh\eta - \eta}\right)^2 \left(\frac{\cosh\eta - 1}{\cosh\theta - 1}\right)^3 - 1, \tag{6}$$

$$\delta_H = \frac{\sinh\theta}{\sinh\eta}\left(\frac{\sinh\theta - \theta}{\sinh\eta - \eta}\right)\left(\frac{\cosh\eta - 1}{\cosh\theta - 1}\right)^2 - 1, \tag{7}$$

$$\eta = \mathrm{Arccosh}\left(\frac{2}{\Omega} - 1\right). \tag{8}$$

Once $\Omega$ of the global universe is specified, $\eta$ is determined by equation (8), and hence $\delta_H$ is related to $\delta$ through the parameter $\theta$ (eqs.[6] and [7]). Although the above results apply for open ($\Omega < 1$ and $\Omega_L < 1$) universes, the corresponding results for closed ($\Omega > 1$ or $\Omega_L > 1$) models are reproduced simply by replacing $\theta \longrightarrow i\theta$ or $\eta \longrightarrow i\eta$. The relation of $\delta_H$ versus $\delta$ is plotted in Figure 1 (solid lines; cf, Figure 1 of Davis et al. 1980).

If $\lambda \equiv \Lambda/(3H^2)$ is not zero, the above procedure cannot be performed analytically. Assuming that $\Lambda$ is the same everywhere, we repeated the calculation numerically for spatially flat ($\Omega + \lambda = 1$) universes. The resulting $\delta$-$\delta_H$ relation is plotted also in Figure 1 (dotted lines; cf, Figure 2 of Peebles 1984). Figure 1 clearly indicates that the $\delta$-$\delta_H$ relation is insensitive to $\Lambda$, especially in the linear ($|\delta| \ll 1$) regime. This is simply because $\Lambda$ acts as a homogeneous density field which hardly affects the dynamics of the perturbed region (Lahav et al. 1991).

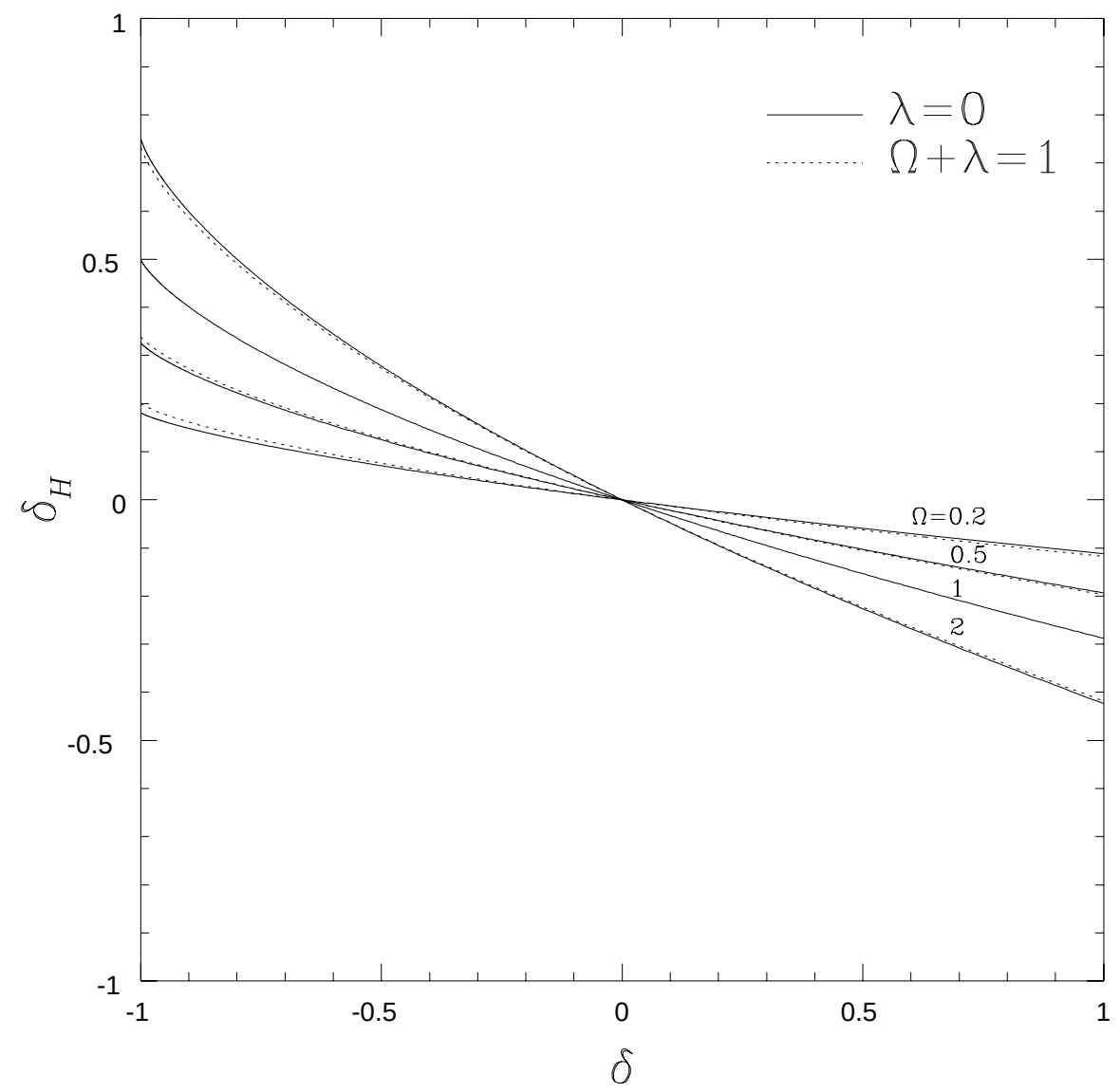

Figure 1: The relation between the density contrast $\delta$ and the "Hubble constant contrast" $\delta_H$ for $\Omega = 0.2, 0.5$, 1 and 2. Solid and dotted curves indicate the relation for $\lambda = 0$ and $\Omega + \lambda = 1$ models, respectively.

For later use, let us define the following quantity $f$ as

$$f(\Omega) \equiv -3\left.\frac{d\delta_H}{d\delta}\right|_{\delta=0} \tag{9}$$

$$= \frac{3(\cosh\eta - 1)(\eta\cosh\eta - 3\sinh\eta + 2\eta)}{\sinh\eta\,(4(\cosh\eta - 1) + \sinh\eta(\sinh\eta - 3\eta))}, \tag{10}$$

where $\eta$ is related to $\Omega$ by equation (8). The second line follows from equations (6) and (7). This quantity is identical to $d\ln D/d\ln a$ where $D$ is the growing solution in linear perturbation theory (e.g., eq.[14.8] in Peebles 1980). Again $f(\Omega, \lambda)$ is very insensitive to $\lambda$; Lahav et al. (1991), for example, approximate $f \simeq \Omega^{0.6} + \frac{1}{70}\lambda(1 + \frac{1}{2}\Omega)$ which is very weakly dependent on $\lambda$. Therefore, we use equations (6), (7) and (10) even when $\Lambda$ is not zero.

## 3. Probability Distribution Function (PDF) of the Global Hubble Constant

In the above spherical model, $\delta$ and $\delta_H$ have one-to-one correspondence. Therefore the conditional PDF of the global Hubble constant $P_{h|\delta}(h|\delta)$ for given $\delta$, is related to that of the local Hubble constant $P_{h_L}(h_L)$ as $P_{h|\delta}(h|\delta)dh = P_{h_L}(h_L)dh_L$. Then the PDF of the global Hubble constant $P_h(h)$ is obtained by the following convolution:

$$P_h(h) = \int P_{h|\delta}(h|\delta)P_\delta(\delta)d\delta$$

$$= \int (1 + \delta_H)P_{h_L}\big(h(1 + \delta_H)\big)P_\delta(\delta)d\delta. \tag{11}$$

Table 1: Model parameters

| Model | $\Omega_0$ | $\lambda_0$ | $h_{obs}$ | $\Delta h_{obs}$ | $R(h^{-1}\mathrm{Mpc})$ | $b$ |
|---|---|---|---|---|---|---|
| E1 | 1 | 0 | 0.8 | 0.08 | 12 | 1 |
| E2 | 1 | 0 | 0.8 | 0.17 | 12 | 1 |
| O | 0.2 | 0 | 0.8 | 0.08 | 12 | 1 |
| L1 | 0.2 | 0.8 | 0.8 | 0.08 | 12 | 1 |
| L2 | 0.2 | 0.8 | 0.8 | 0.17 | 12 | 1 |
| L3 | 0.2 | 0.8 | 0.8 | 0.17 | 70 | 1 |
| TCO | 1 | 0 | 0.8 | 0.08 | 30 | 1.3 |

Since $P_{h_L}(h_L)$ is unknown a priori, we adopt a natural and reasonable choice of Gaussian:

$$P_{h_L}(h_L) = \frac{1}{\sqrt{2\pi}\Delta h_{obs}} \exp\left[-\frac{(h_L - h_{obs})^2}{2\Delta h_{obs}^2}\right]. \tag{12}$$

At present, the uncertainty of the observation $\Delta h_{obs}$ is dominated by systematic errors (see Table 1 of Freedman et al. 1994). As the number of the sample galaxies increases, however, statistical errors will dominate and the assumption of Gaussian will be more realistic. According to the widely accepted view, the density fluctuation field is assumed to be random-Gaussian with the rms value $\sigma$:

$$P_\delta(\delta) = \frac{1}{\sqrt{2\pi}\sigma} \exp\left[-\frac{\delta^2}{2\sigma^2}\right]. \tag{13}$$

Equation (11) together with equations (6), (7), (12) and (13) yields the analytical expression for the PDF of the global Hubble constant. TCO computed this PDF numerically by the similar convolution assuming that $P_{\delta_H|h}(\delta_H|h)$ is the same as $P_{\delta_H|h_L}(\delta_H|h_L)$.

The value of $\sigma$ is computed from the power spectrum $P(k)$ of the density fluctuation:

$$\sigma^2(R) = \frac{1}{2\pi^2}\int_0^\infty P(k)W^2(kR)k^2dk, \qquad W(x) \equiv \frac{3}{x^3}(\sin x - x\cos x), \tag{14}$$

where we assumed the top-hat window function $W(x)$ with $R$ being the smoothing length. In what follows, we consider the CDM models for specific examples, and adopt the fit by Bardeen et al. (1986) for $P(k)$. The value of $\sigma$ is normalized to be $b^{-1}$ at $R = 8h^{-1}$Mpc ($b$ is the biasing parameter). Note, however, that the above result depends on cosmological scenarios only through $\sigma$ (via $P(k)$). Thus our conclusions below apply for any scenarios as long as $\sigma$ takes the same value.

Our model parameters are summarized in Table 1. We adopt a set of parameters used by TCO for comparison. We consider the smoothing length $R = 12h^{-1}$Mpc and $70h^{-1}$Mpc corresponding to the recession velocities of the Virgo ($= 1180 \pm 22$ km/s, Sandage & Tamman 1990) and of the Coma ($= 6931 \pm 45$ km/s, Aaronson et al. 1986), respectively. We have numerically integrated the PDF (11) with equations (6), (7), (12), (13) and (14), and the results are plotted in Figure 2a. In practice, the integration is performed over $-1 < \delta < \infty$ (viz., $\infty \to 0$ and $0 \to 2\pi i$ for $\theta$). For comparison, the Gaussian distribution of $P_{h_L}$ with $(h_{obs}, \Delta h_{obs}) = (0.8, 0.08)$ is shown in dotted curve; the Gaussian curve with $(0.8, 0.17)$ is almost identical to the model L3, because $\sigma$ is so small that the probability distribution is dominated by the observational error (see eq.[17]). The peaks of $P_h(h)$ are slightly shifted to the left of $h_{obs}$, especially in models E1 and E2. This shift is originated from the convolution process and can be understood by differentiating equation (11) with respect to $h$.

Since we do not take into account the negligibly small $\lambda$-dependence in the $\delta$-$\delta_H$ relation (Figure 1), the results for models O and L1 are identical. To see the parameter dependence of the shape of the PDF, we approximate equation (11) using linear theory, i.e., $\delta_H = -f\delta/3$, where $f$ is defined in equation (10):

$$P_h(h) \simeq \frac{1}{\sqrt{2\pi}} \frac{\Delta h_{obs}^2 + f^2\sigma^2 h_{obs} h/9}{(\Delta h_{obs}^2 + f^2\sigma^2 h^2/9)^{3/2}} \exp\left[-\frac{1}{2}\frac{(h-h_{obs})^2}{\Delta h_{obs}^2 + f^2\sigma^2 h^2/9}\right], \tag{15}$$

where integration was carried out formally from $-\infty$ to $\infty$ with respect to $\delta$. This linear approximation is justified only when $\sigma$ is much less than unity and when the integration in equation (11) is contributed mainly from the linear regime. Nevertheless equation (15) provides a good insight into the general feature of $P_h(h)$. Equation (15) indicates that $P_h(h)$ is not symmetric around $h_{obs}$ and the tail to the right falls off less rapidly than to the left. Hence the expectation value $\langle h \rangle$:

$$\langle h \rangle \equiv \int_0^\infty h P_h(h) dh \Big/ \int_0^\infty P_h(h) dh \tag{16}$$

becomes slightly greater than $h_{obs}$ even though the peak position is smaller than $h_{obs}$. It should be noted that $P_h(h)$ in the present case is not normalized in an exact sense; the Gaussian distribution (13) assumes that $\delta$ ranges from $-\infty$ to $\infty$ despite that $\delta \geq -1$ in reality. In addition, we consider the expanding ($h > 0$) models only. For these two reasons, we define $\langle h \rangle$ by equation (16). The standard deviation is very roughly approximated as

$$\Delta h \equiv \sqrt{\langle (h - \langle h \rangle)^2 \rangle} \sim \sqrt{\Delta h_{obs}^2 + f^2\sigma^2 h_{obs}^2/9} \simeq \Delta h_{obs} + \frac{\Omega_0^{1.2}\sigma^2 h_{obs}^2}{18\Delta h_{obs}}, \tag{17}$$

implying that the width $\Delta h_{obs}$ of the original PDF is broadened to $\Delta h$ by the convolution. The effect of the convolution $\Delta h - \Delta h_{obs}$ increases monotonically with $\Omega_0$ as seen in Figure 2a, and decreases with $R$ and $b$ simply because $\sigma$ is generally a decreasing function of $R$ and $b$. In Figure 2b is plotted the cumulative probability $P_{h<}(h)$ that the Hubble constant is less than $h$, defined by

$$P_{h<}(h) \equiv \int_0^h P_h(h) dh \Big/ \int_0^\infty P_h(h) dh. \tag{18}$$

Similarly one can derive the PDF of the age of the universe $t_0$ or $\tau \equiv t_0/\mathrm{Gyr}$ given the local estimate of $H_0$, through $P_\tau(\tau) d\tau = P_h(h) dh$. Combining with

$$t_0 = \frac{T}{H_0}, \qquad T(\Omega_0, \lambda_0) \equiv \int_0^1 \left[\frac{\Omega_0}{x} + (1 - \Omega_0 - \lambda_0) + \lambda_0 x^2\right]^{-1/2} dx, \tag{19}$$

one finds

$$P_\tau(\tau) = \frac{\alpha T}{\tau^2} P_h\left(\frac{\alpha T}{\tau}\right), \qquad \alpha \equiv (100 \text{ km/s/Mpc} \times \text{Gyr})^{-1} = 9.78\cdots, \tag{20}$$

which is plotted in Figure 3a. Since we neglect the effect of $\lambda_0$ on $P_h(h)$, $P_\tau(\tau)$ depends on $\lambda_0$ only through $T = T(\Omega_0, \lambda_0)$. Although $T$ decreases with $\Omega_0$, we numerically found that $\Delta\tau \equiv \sqrt{\langle (\tau - \langle \tau \rangle)^2 \rangle} = \alpha T \sqrt{\langle h^{-2} \rangle - \langle h^{-1} \rangle^2}$ is an increasing function of $\Omega_0$. Nevertheless $\langle \tau \rangle + \Delta\tau$ still decreases with $\Omega_0$. In Figure 3b is plotted the cumulative probability $P_{\tau>}(\tau) \equiv P_{h<}(\alpha T/\tau)$ that the age of the universe is older than $\tau$ Gyr.

Our predictions on the models are summarized in Table 2, which list the expectation values of $h$ and $\tau$ with $1\sigma$ error, the probability that the Hubble constant is less than 50 km/s/Mpc, and the probability that the age of the universe is older than 14 Gyr.

## 4. Discussion and Conclusion

We have presented a method to compute the probability distribution function of the (true) Hubble constant and the age of the universe, given the estimate of the Hubble constant in our nearby galaxy samples. Our method takes into account both the observational errors and the cosmic variance, and enables to quantitatively compute the constraints on the cosmological models.

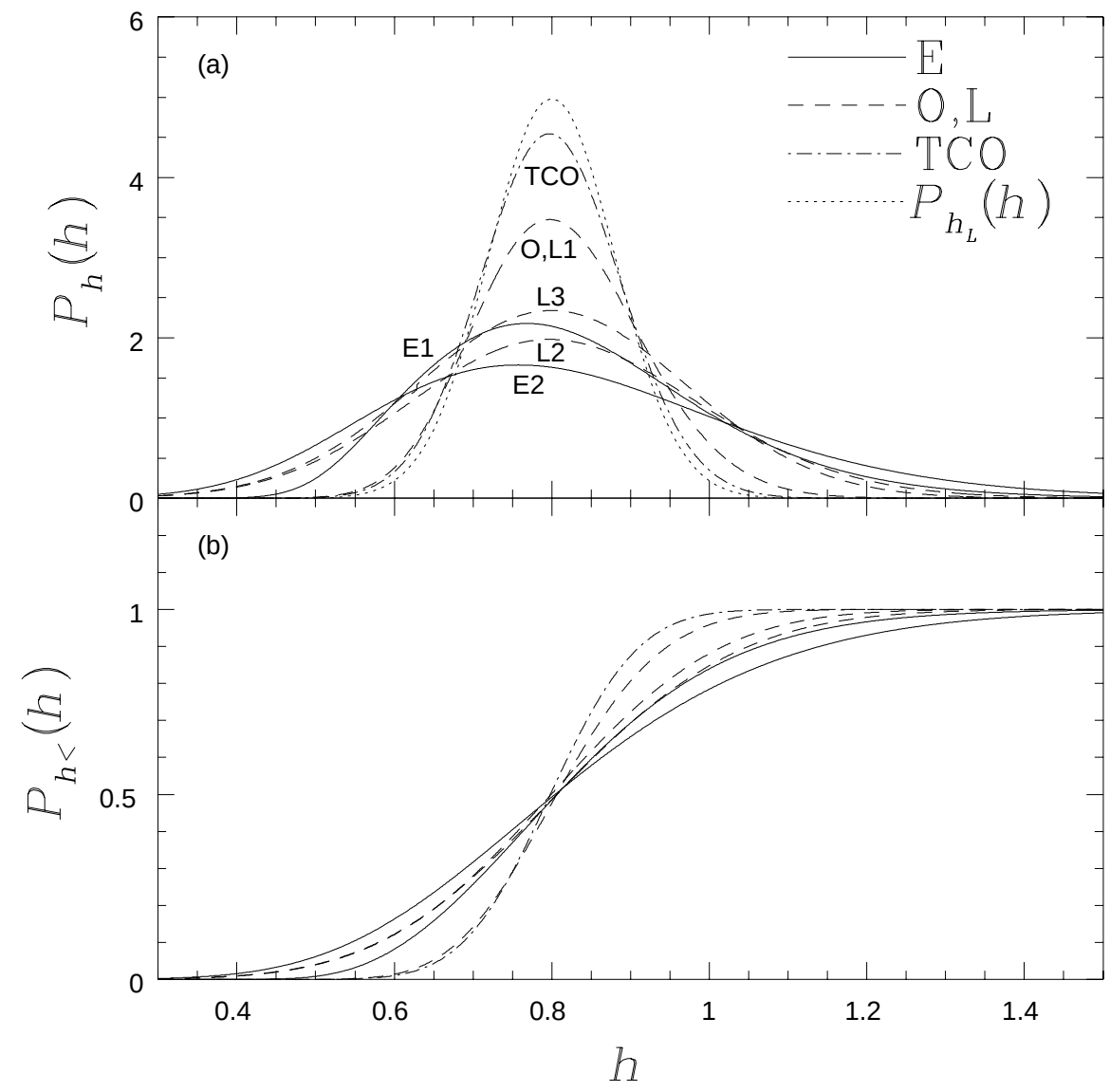

Figure 2: a) Probability distribution of the Hubble constant $P_h(h)$ (eq.[11]); b) Cumulative probability $P_{h<}(h)$ that the Hubble constant is less than $h$ (eq.[18]).

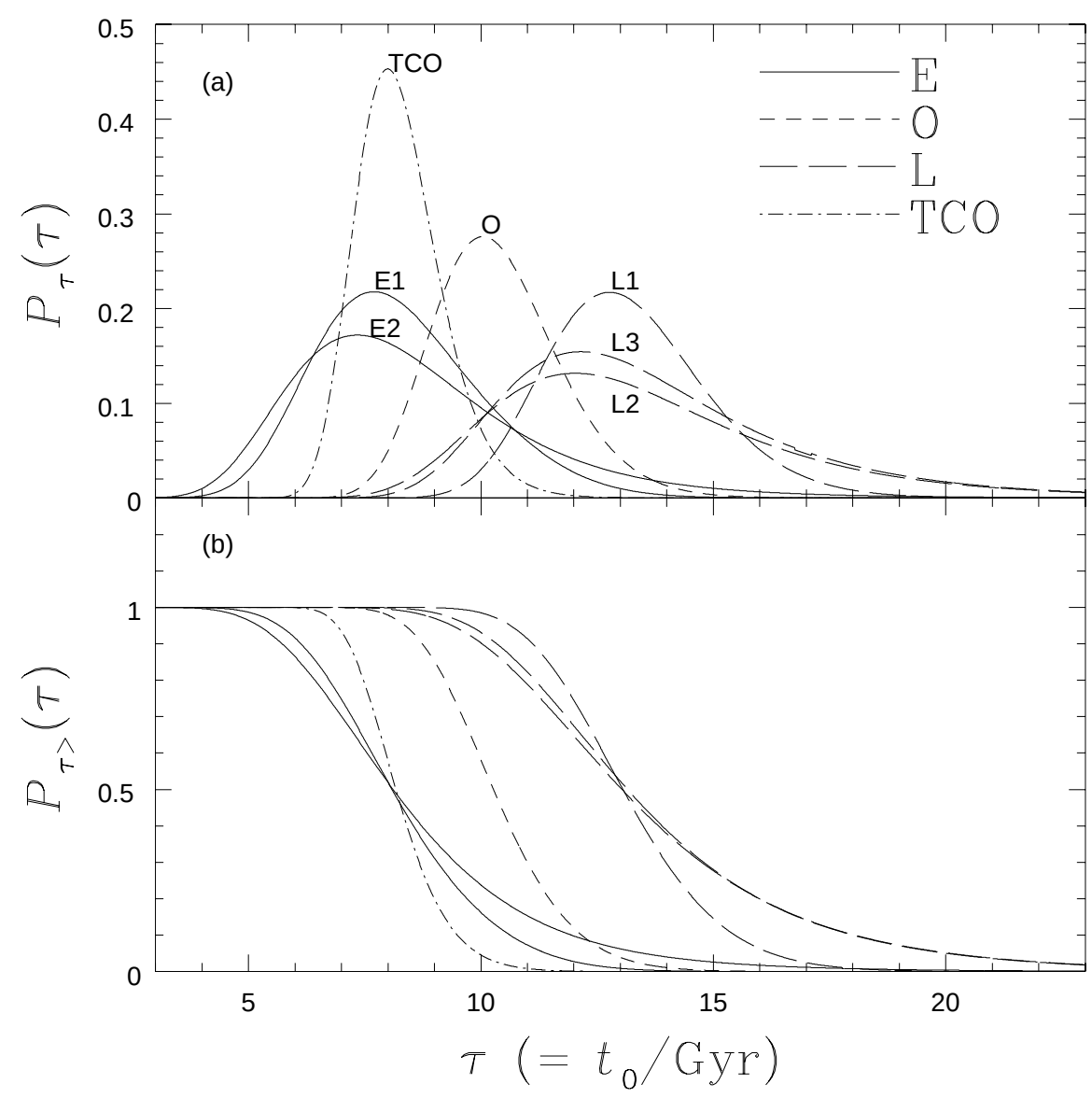

Figure 3: a) Probability distribution of the age of the universe $P_\tau(\tau)$ (eq.[20]); b) Cumulative probability $P_{\tau>}(\tau)$ that the age of the universe is older than $\tau$ Gyr.

Table 2: Summary of the prediction on the CDM models

| Model | $\langle h\rangle \pm \Delta h$ | $\langle \tau\rangle \pm \Delta\tau$ | $P_{h<}(0.5)$ | $P_{\tau>}(14)$ |
|---|---|---|---|---|
| E1 | 0.83±0.18 | 8.3±1.8 | $8.4\times10^{-3}$ | $2.5\times10^{-3}$ |
| E2 | 0.83±0.24 | 8.6±2.9 | $6.1\times10^{-2}$ | $4.1\times10^{-2}$ |
| O | 0.81±0.10 | 10.4±1.4 | $3.2\times10^{-4}$ | $1.1\times10^{-2}$ |
| L1 | 0.81±0.10 | 13.2±1.7 | $3.2\times10^{-4}$ | $3.0\times10^{-1}$ |
| L2 | 0.81±0.18 | 13.7±4.0 | $4.0\times10^{-2}$ | $3.8\times10^{-1}$ |
| L3 | 0.80±0.17 | 13.9±3.9 | $4.0\times10^{-2}$ | $3.9\times10^{-1}$ |
| TCO | 0.80±0.09 | 8.3±0.9 | $2.2\times10^{-4}$ | $4.2\times10^{-5}$ |

Before summarizing the implications of our results, let us discuss the validity of the spherically symmetric approximation which we assumed throughout the present *Letter*. We described the dynamics of the local universe centered around us using the spherical nonlinear model with the mean density contrast smoothed over the same effective volume. Since the scales of our interest here are still in linear regime, this approximation is statistically valid in practice. To see this in more detail, let us compare our $P_h(h)$ with that computed numerically by TCO. Specifically the model TCO in our Figure 2a should be compared with their Figure 7. Qualitatively they are in agreement, but more careful examination reveals that our PDF is more strongly peaked around $h = h_{obs}$. In fact we believe that this difference is explained by the fact that they approximated the PDF for $\delta_H$ by the Gaussian; as is clear from their Figure 1a, the PDF computed directly from the numerical simulations is more strongly peaked than the Gaussian fit especially on small scales. In fact our PDF $P_{h|\delta}(h|\delta)$ properly takes account of such departure from the Gaussian distribution. Therefore the quantitative difference is not due to the assumption of the spherical symmetry. Also it is interesting to note that our model E1 is quite similar to their result based only on data from octant of the sky. TCO ascribed the broadening of the PDF to the angular variance in the expansion rate, but Figure 2a implies that most of the effect can be explained simply by larger $\sigma$ in small sampling volume (cf, eq.[17]); the smoothing scale $R = 12h^{-1}$Mpc in our model E1 is roughly the same as their *effective* sampling radius 30/2 $h^{-1}$Mpc. Thus we conclude that our assumption of spherical symmetry in the present context is justified on $R \gtrsim 10\ h^{-1}$Mpc at least in CDM models.

Applying our method of computing the PDF to the CDM models, one can discuss quantitatively the confidence level of the (global) cosmological parameters inferred from the local observations. Let us assume, for instance, that the age of the present universe is larger than 14 Gyr. Then as shown in Table 2, the Einstein – de Sitter models are rejected with (96 $\sim$ 99.7) % level (models E2 and E1) even if we allow for the relatively big quoted error in the current local observation. Similarly the model O ($\Omega_0 = 0.2$ and $\lambda_0 = 0$) is viable only with 1 % confidence level. Naturally non-zero $\lambda$ models are strongly favored.

Currently the observational error of the Hubble constant is dominated by the back-to-front geometry of the Virgo cluster itself (Freedman et al. 1994), and therefore we do not have strong constraints on the non-zero $\lambda$ models. As the number of sample galaxies increases, however, such a systematic error will be reduced significantly. Then our probabilistic approach will provide a powerful tool in quantitatively constraining the (global) cosmological parameters.

This research was supported in part by the Grants-in-Aid by the Ministry of Education, Science and Culture of Japan (05640312, 06233209).